\documentclass{sig-alternate}

\usepackage[utf8]{inputenc}
\usepackage{amsmath}
\usepackage{amsfonts}
\usepackage{mathtools}
\usepackage{xcolor}
\usepackage{booktabs}

\makeatletter
\newif\if@restonecol
\makeatother

\usepackage[ruled,lined,linesnumbered]{algorithm2e}

\def \calG {\mathcal{G}}

\def \calE {\mathcal{E}}

\makeatletter
\def\@copyrightspace{\relax}
\makeatother

\begin{document}

\title{Social Events in a Time-Varying Mobile Phone Graph}

\numberofauthors{6}

\author{
\alignauthor
Carlos Sarraute\\
       \affaddr{Grandata Labs}\\
       \texttt{charles@grandata.com}
\alignauthor
Jorge Brea\\
       \affaddr{Grandata Labs}\\
       \texttt{jorge@grandata.com}
\alignauthor
Javier Burroni\\
       \affaddr{Grandata Labs}\\
       \texttt{javier.burroni@grandata.com}
\and 
\alignauthor
Klaus Wehmuth\\
       \affaddr{LNCC}\\
       \texttt{klaus@lncc.br}
\alignauthor
Artur Ziviani\\
       \affaddr{LNCC}\\
       \texttt{ziviani@lncc.br}
\alignauthor
J.I. Alvarez-Hamelin\\
       \affaddr{UBA--CONICET}\\
       \texttt{ihameli@cnet.fi.uba.ar}
}

\date{}
\maketitle

\section*{Introduction}

The large-scale study of human mobility has been significantly enhanced over the last decade by the massive use of mobile phones in urban populations~\cite{Becker2013,Calabrese:2014}. Studying the activity of mobile phones allows us, not only to infer social networks between individuals, but also to observe the movements of these individuals in space and time~\cite{ponieman2013human}. 
In this work, we investigate how these two related sources of information can be integrated within the context of detecting and analyzing large social events. We show that large social events can be characterized not only by an anomalous increase in activity of the antennas in the neighborhood of the event, but also by an increase in social relationships of the attendants present in the event. Moreover, having detected a large social event via increased antenna activity, we can use the network connections to infer whether an unobserved user 
was present at the event.  More precisely, we address the following three challenges: 
(i)~automatically detecting large social events via increased antenna activity; 
(ii)~characterizing the social cohesion of the detected event;  
and (iii)~analyzing the feasibility of inferring whether unobserved users were in the event.

\section*{Data Source and Methodology} \label{sec:data-source}

Our data source is an anonymized traffic dataset from a mobile phone operator in Argentina, collected mostly in the Buenos Aires metropolitan area, over a period of 5 months. 
The raw data logs contain about 50 million calls per day.
Call Detail Records~(CDR) are an attractive source of location information since they are collected 
for all active cellular users~(about 40 million users in Argentina).
Further, additional uses of CDR data incur little marginal cost. 

Together with the location of the clients of the mobile phone network, the CDRs allow us to reconstruct a social graph derived from the communications among users. 
We first define a contact graph $\calG$ composed of all users participating in the mobile network, where an edge between users exist if they have communicated at any point within the 5~months period. 
We note that $\calG$ includes non client users that have communicated with client users. 

Then, we define a time series of subgraphs of $\calG$ composed only of users present at a given location at a given time. In this work, these subgraphs are restricted to client nodes (users for which we have mobility data) using the event antenna at a specified day, for a given time window. Note that the existence of a link between such nodes in any subgraph depends on whether there is a link between them in~$\calG$. These subgraphs allow us to focus on the social dynamics of a set of users present at a particular location at a given time as is the case of client users attending a large-scale event.

We formalize this temporal sequence of subgraphs within the context of a Time-Varying Graph (TVG) based on the representation encountered in~\cite{Wehmuth2014}. 
We consider a TVG as an object $H = (V,E,T)$, where $V$ is a vertex set, $T$ is a time instant set, and $E$ is an edge set. In this context, an edge $e \in E$ is primarily a quadruple of the form $\left<u, t_a, v, t_b \right>$, where $u,v \in V$ are vertices and $t_a, t_b \in T$ are time instants. In this representation, an edge $\left<u, t_a, v, t_b \right>$ then shows a relation between vertex $u$ at time $t_a$ and vertex $v$ at time $t_b$, i.e. a directed relation, where the pair $\left< u, t_a \right>$ is the origin and the pair $\left< v, t_b \right>$ is the destination.  Note that the represented relation is between the composite objects $\mathbf{u} = \left< u, t_a \right>$ and $\mathbf{v} = \left< v, t_b \right>$.  In this representation, a TVG edge $e \in E \subset  V \times T \times V \times T$ can be thus seen as an ordered pair $\left< \mathbf{u}, \mathbf{v} \right> \in (V \times T) \times (V \times T)$ of composite vertices.

In this work, each edge in a subgraph is represented as a tuple $\left < (x,t_x),(y,t_y) \right >$, where user $x$ is the caller, user $y$ is the callee, $t_x=t_y$ is the date and time of the call. In addition to the origin and destination composite vertices present on the TVG edge, we also wish to identify the local base station antenna~(BSA) used on the call represented by each TVG edge. In order to achieve this, we represent the BSA identity as an edge attribute, so that the TVG edge used in this particular work becomes an ordered 5-tuple of the format $\left< x, t_x, y, t_y, l \right>$ where in addition to the known edge elements, $l$ represents the BSA identity.

\section*{Automatic Event Detection} \label{sec:automatic-event-detection}

In this section, we show how we can use the mobile phone data in an endogenous manner to detect past events automatically (as \cite{chua2013automatic,ritter2012open} did with social media such as Twitter).
The events we are interested in involve a large amount of people, hence they significantly increase the usage demand of antennas nearby, as evidenced by their traffic time series.
Our strategy thus is to locate time windows where the antenna activity is significantly higher than some appropriate baseline. Defining an appropriate activity baseline is not a trivial task since this is possibly dependent on the size of the event one is trying to detect and on the regularity of the normal antenna activity patterns. 

\begin{figure}[bh]
  \centering
  \includegraphics[width=0.45\textwidth, trim=1.5cm 0cm 1.5cm 1.0cm, clip=true]{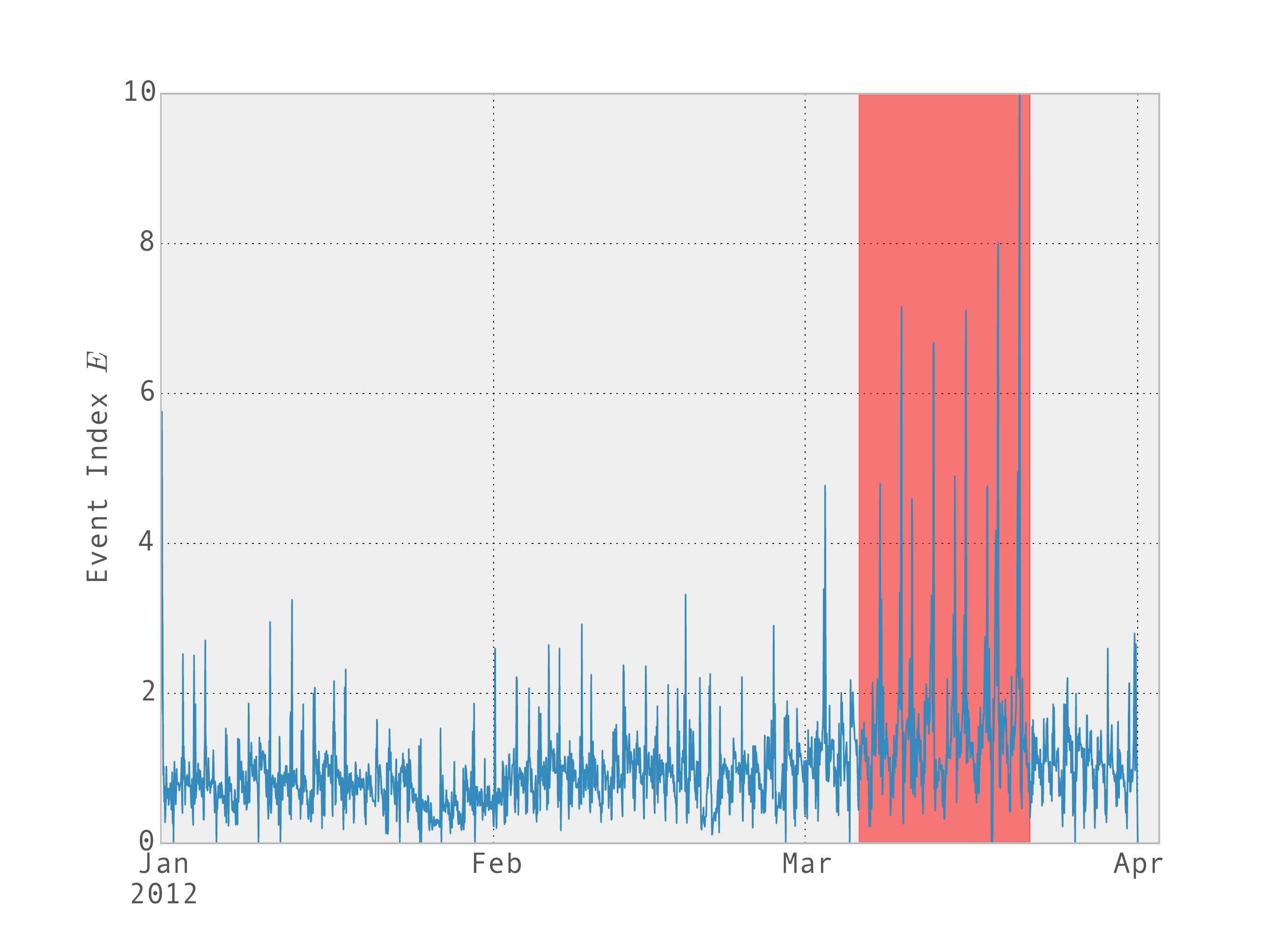}
  \caption{Event index $\calE$ for an antenna near the River Plate soccer stadium. 
  }
  \label{fig:roger_waters}
\end{figure}

In this work, the antennas activity is given by the number of calls per hour registered in each antenna, according to the week $w_i$ ($0 \leq i < n$, where $n$ is the number of weeks in the studied dataset), the day of the week $d_j$ ($ 0 \leq j \leq 6$), 
and the hour of the day $h_k$ ($0 \leq k \leq 23$). 
For each time slot $(w_i, d_j, h_k)$, we denote the number of calls as $ C_{(i,j,k)} $ which we then normalise be the average activity over a period of 3 months,
defining an \emph{event index} $\calE_{(i,j,k)} $ given by
\[
\label{eq:event_index}
\calE_{(i,j,k)} = \frac{ C_{(i,j,k)} } { \frac{1}{n} \sum_{\ell} C_{(\ell,j,k)}  } 
\]
for the $i^{th}$ week in the 3 month period, $j^{th}$ day of the week, $k^{th}$ hour of the day, and where $\ell$ sums over all weeks in the whole time period considered ($0 \leq \ell < n$).
Figure~\ref{fig:roger_waters} shows the event index $\calE$
for an antenna near the stadium of River Plate, a major soccer team in Buenos Aires, 
for each time slot (hour) from January 1 to March 31, 2012. 
We consider a large-scale event to have occurred when the event index is within the highest $99$-percentile while considering the whole 3 months period.

During January and February, we observe the event index smoothly oscillating around the mean value
except for small sharp increases due to normal peak daily activity. As we approach March, the oscillations' amplitude starts to increase reaching its peak value by mid March. On top of this large oscillations we see very large sharp peaks reaching values of $\calE=9.98$ on March 20. Additional high peaks are also observed on March 7, 9, 10, 12, 14, 15, 17, and 18. Retrospectively, we corroborated that a sequence of 9 Roger Waters concerts took place at this location during March on the same days of the observed peaks. We thus highlight that the event index has clearly spotted the Roger Waters events on the exact days the concerts took place. 

An open question is to understand the slight increase in the background oscillation during these peak periods. A plausible explanation is an increase in activity due to gatherings before the event (e.g., fans trying to get tickets, organizers working at the stadium, January and February are vacations month in Argentina, etc.) or it might be a statistical fluctuation one could verify if one had access to a much longer dataset. In any case, further study of this background activity can prove very useful, for instance, allowing for the forecasting of large-scale events. Identifying these early precursors  could allow us not only to automatically detect events retrospectively, but also give us the possibility of anticipating a future event.

\section*{Social Dimension of Events}

Having detected a large-scale event, we now take a look at social relationships of users attending these events. We consider all communications from the event antenna for a period of 5~months from November 1, 2011 to March 31, 2012, therefore including the Roger Waters concert series from March 7--20, 2012. For each day, we compute the implicit social subgraph induced by $\calG$ restricted to users observed at the event antenna during the time window from 18:00 to 22:00 (time window for the Roger Waters concerts).

\begin{figure}[th]
\centering
\begin{footnotesize}
	\begin{minipage}{.24\linewidth}
		\centering
		\includegraphics[width=0.9\textwidth]{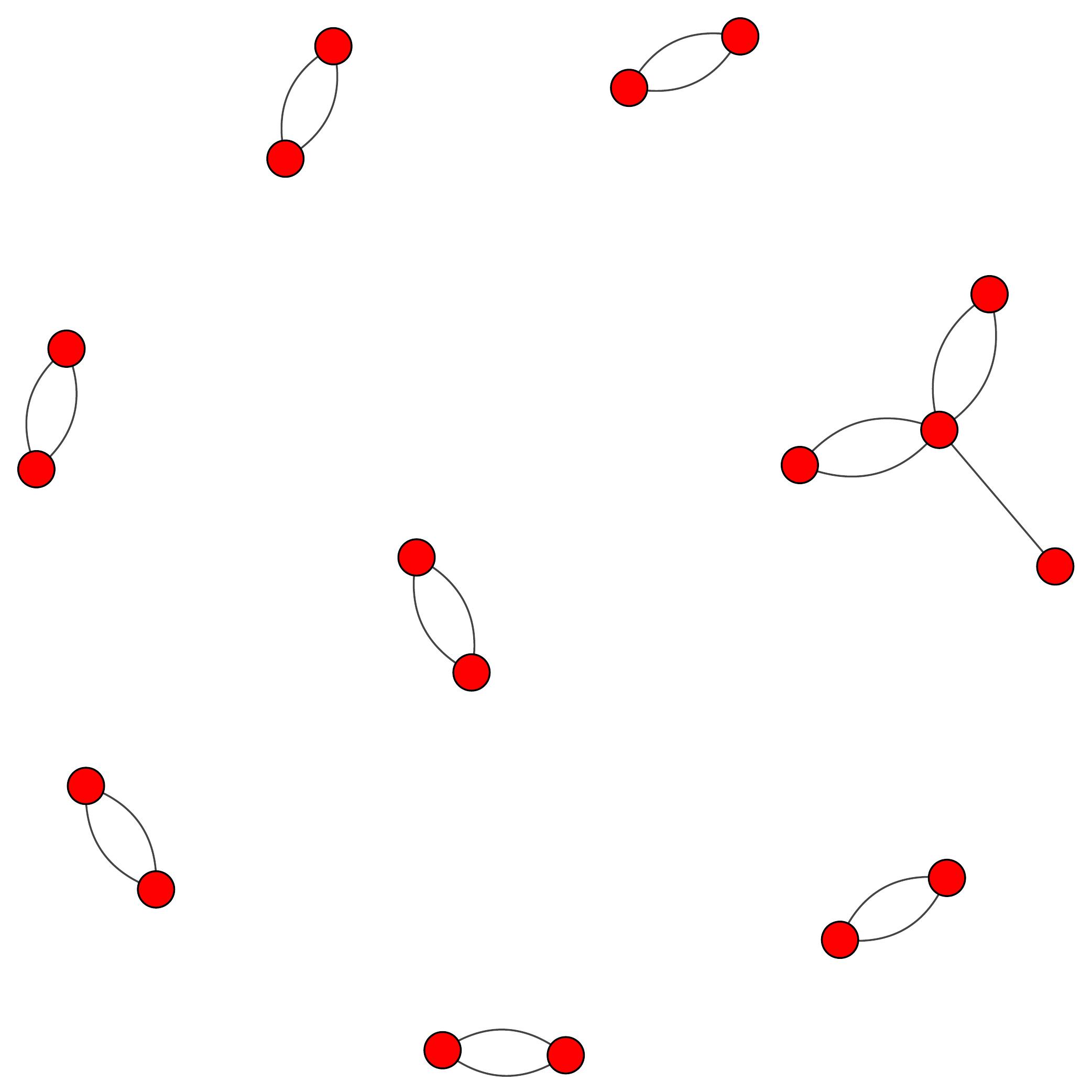} 
		\small{(a) Feb 7} 
	\end{minipage}
	\begin{minipage}{.24\linewidth}
		\centering
		\includegraphics[width=0.9\textwidth]{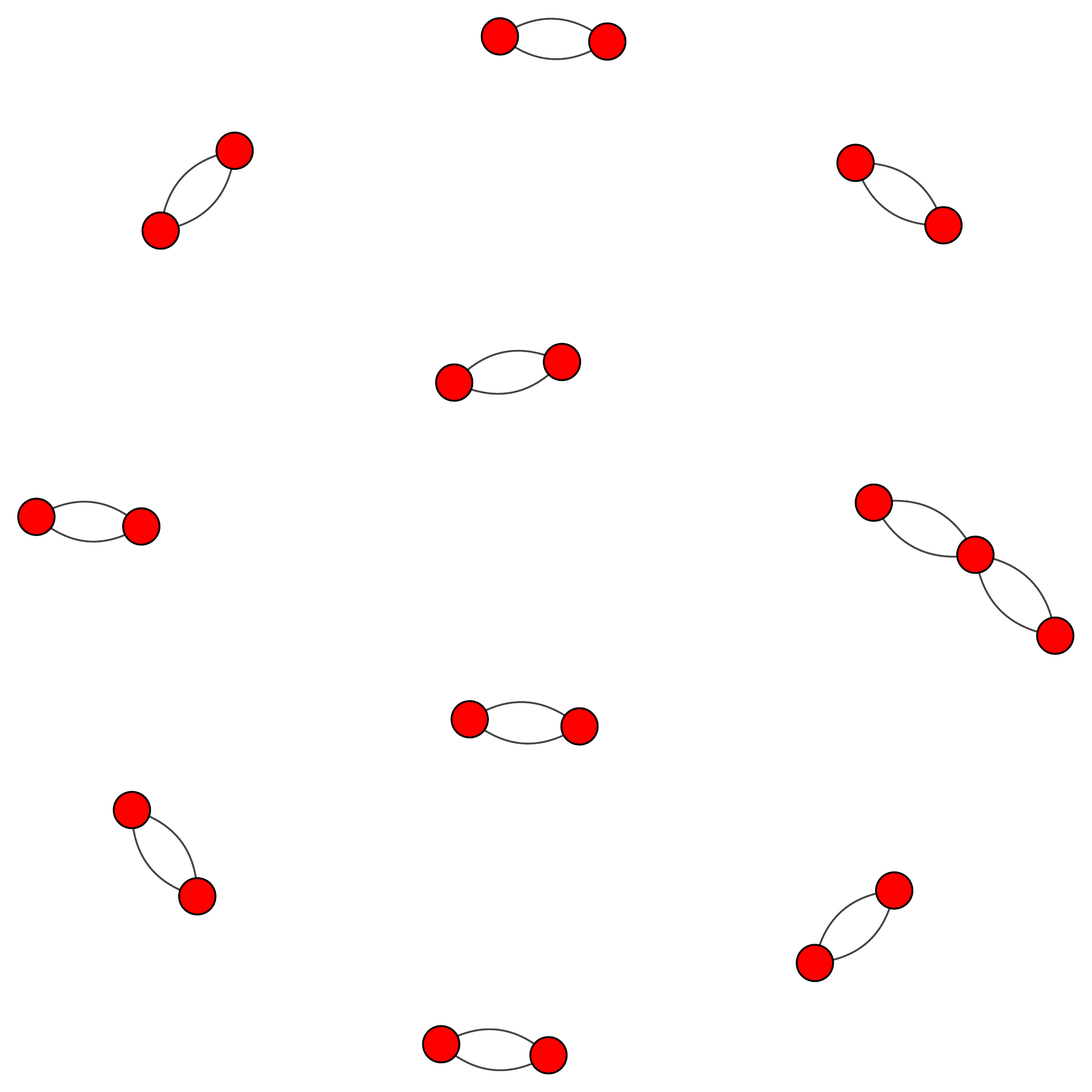}
		\small{(b) Feb 17} 
	\end{minipage}
	\begin{minipage}{.24\linewidth}
		\centering
		\includegraphics[width=0.9\textwidth]{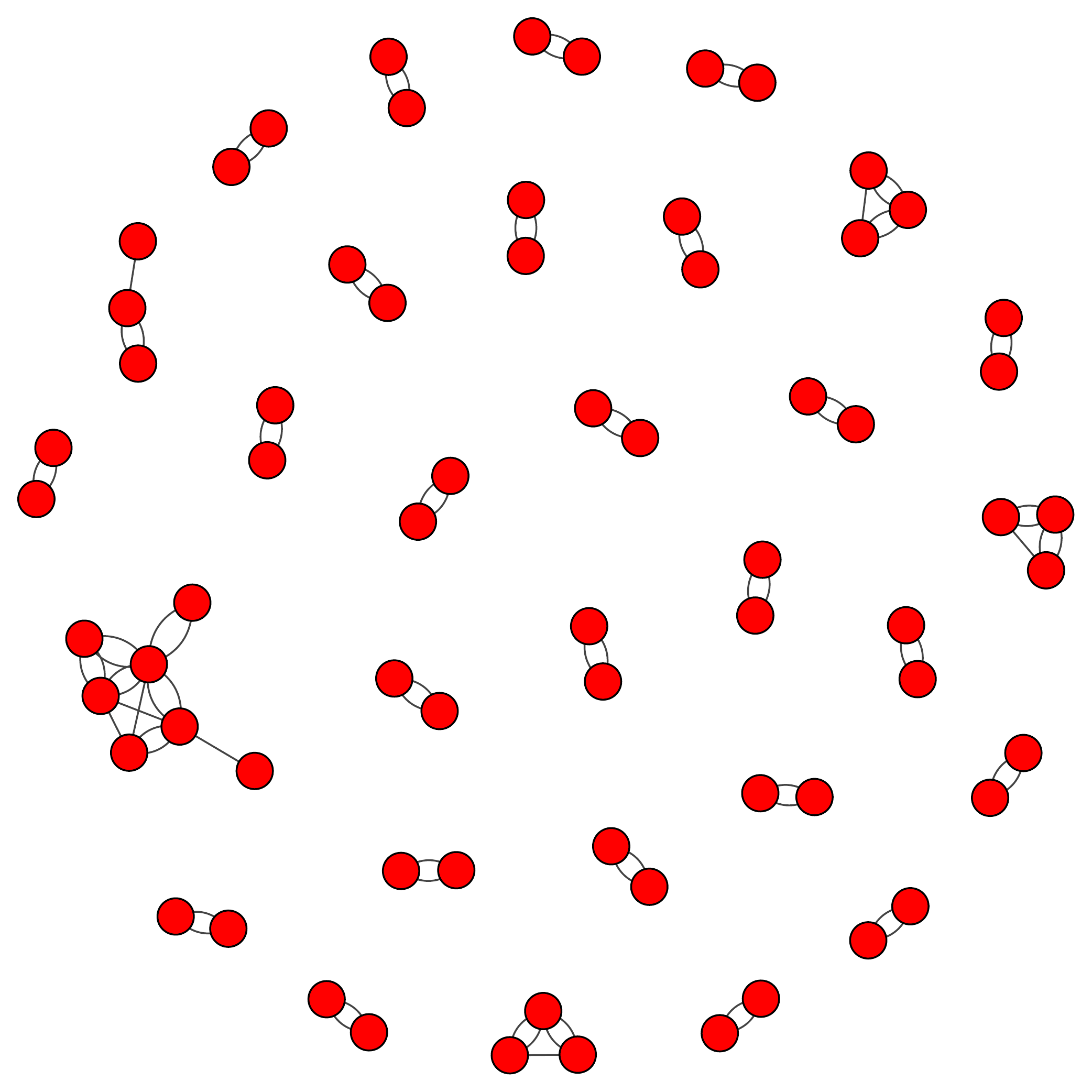}
		\small{(c) March 7} 
	\end{minipage}
	\begin{minipage}{.24\linewidth}
		\centering
		\includegraphics[width=0.9\textwidth]{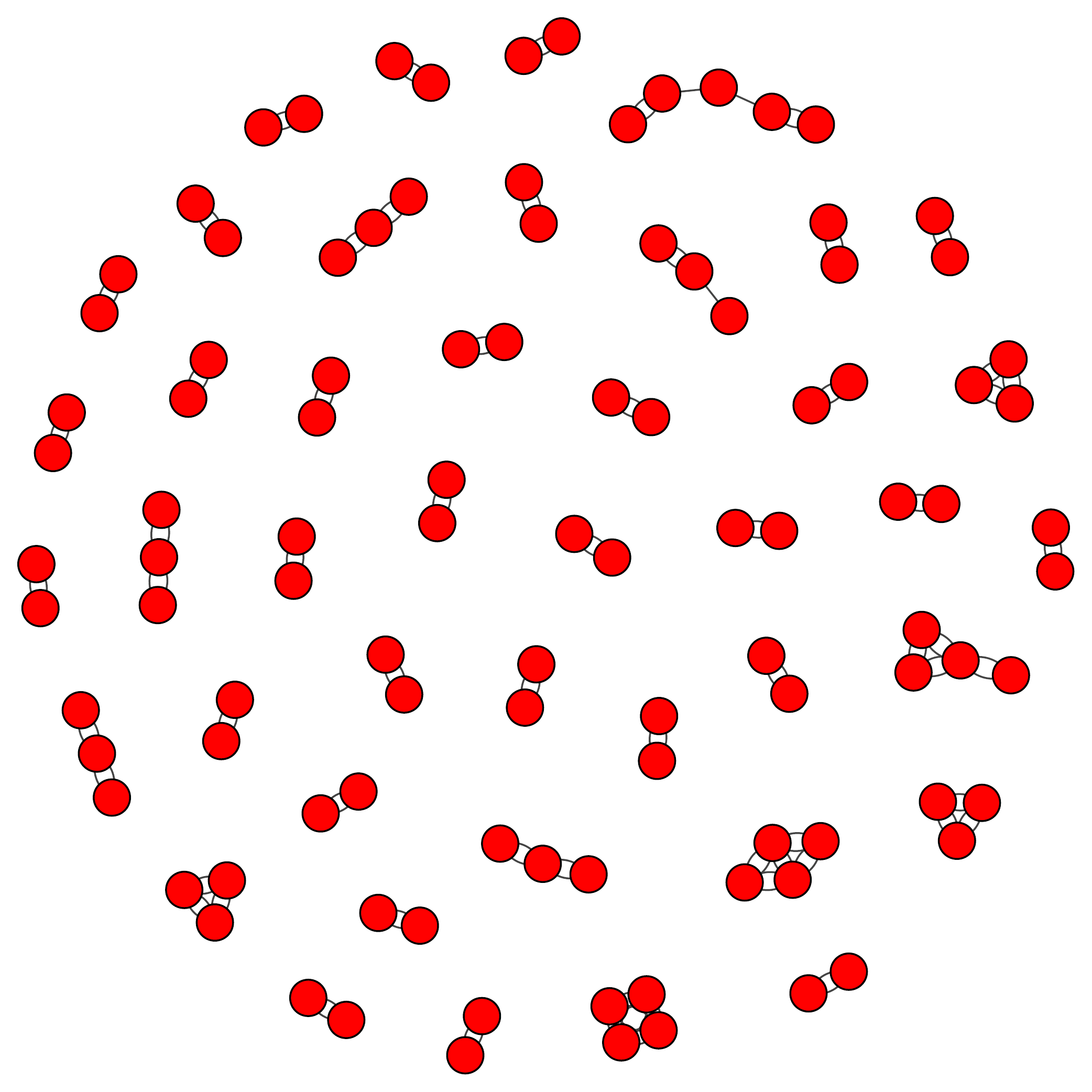}
		\small{(d) March 17} 
	\end{minipage}
\end{footnotesize}
\vspace{-0.2cm}
\caption{Implicit social graphs: (a), (b) two days with no large event; (c), (d) two days of the Roger Waters concerts. 
}
\label{SocialCohesionEvents}

  \label{fig:social_cohesion_non_event}
\end{figure}

Figure \ref{SocialCohesionEvents} shows the induced subgraphs for February 7 and 17, in which no event took place 
(a, b),
along with March 7 and 17 for which two Roger Waters concerts took place 
(c, d).
The total number of nodes on February 7 and 17 were 107 and 130, respectively, whereas on March 7 and 17 we observe 716 and 740 attenders, respectively. Most nodes are singlets and are thus not shown in the plots, leaving only nodes for whom at least one of his/her contacts in $\calG$ was also present at the event location, at the same day, and during the defined time window. 

Therefore, one can expect that it is somewhat likely that the nodes present in these subgraphs are attending this location together. We call these nodes \emph{social attenders} and all nodes, singlets included, \emph{attenders}. An immediate observation is the larger number of social attenders at the days of the concerts than on a normal day. This is expected given the larger attendance during the event. More interestingly, for the days of the event, especially on March 17, we observe more complex group relationships of up to 7 nodes, possibly indicating larger groups of people attending the concert together.

\subsection*{Probability of Being in an Event}

\begin{figure}[ht]	
\begin{center}
	\includegraphics[width=0.45\textwidth]{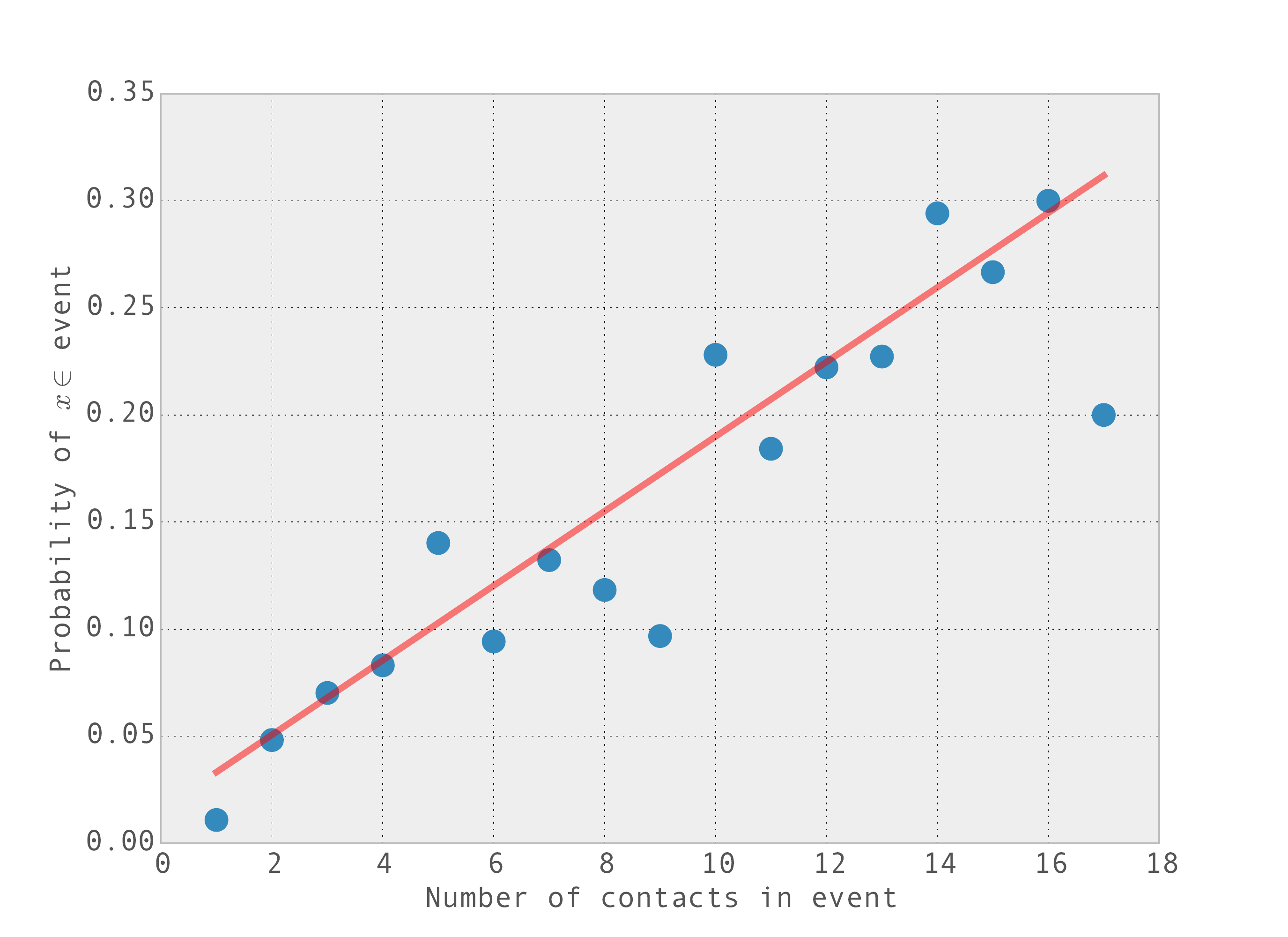}
\end{center}
\caption{Scatter plot and linear regression curve for the conditional 
probability of a user being in an event given that $k$ of his/her contacts in graph $\calG$ were in the event. 
}
\label{fig:probability}
\end{figure}

The increased social structure suggested in Fig.~\ref{SocialCohesionEvents} motivates the following question:
can we estimate the probability that a user attended the event, given that $k$ of his/her contacts in graph $\calG$ also attended the event? We compute this  probability as follows:
\[
	p(x \in U| k \text{ contacts} \in U) = \frac{\text{users} \in U \text{ with } k \text{ contacts} \in U }{\text{users with } k \text{ contacts} \in U},
\]
where $U$ is the set of all users in the event. 
The results are shown in Fig.~\ref{fig:probability}, where we also plot a linear regression curve showing how strongly the expectation of finding a user in the event depends on the amount of graph contacts this user has in the event. The scatter plot and the regression curve ($R=0.83$) show that the probability that a given user attended the event grows linearly with the number of his/her contacts in the event. 
If several of the user contacts in $\calG$ represent real social relationships, such as friends or family members, and a significant fraction of such contacts are in a 
social event, such as a music concert, one could expect an increase in the likelihood the user is also attending the event. Nevertheless, it is worth noticing that the linear relation does no saturate even for high values of~$k$.

\begin{figure}[ht]	
\begin{center}
	\includegraphics[width=0.45\textwidth]{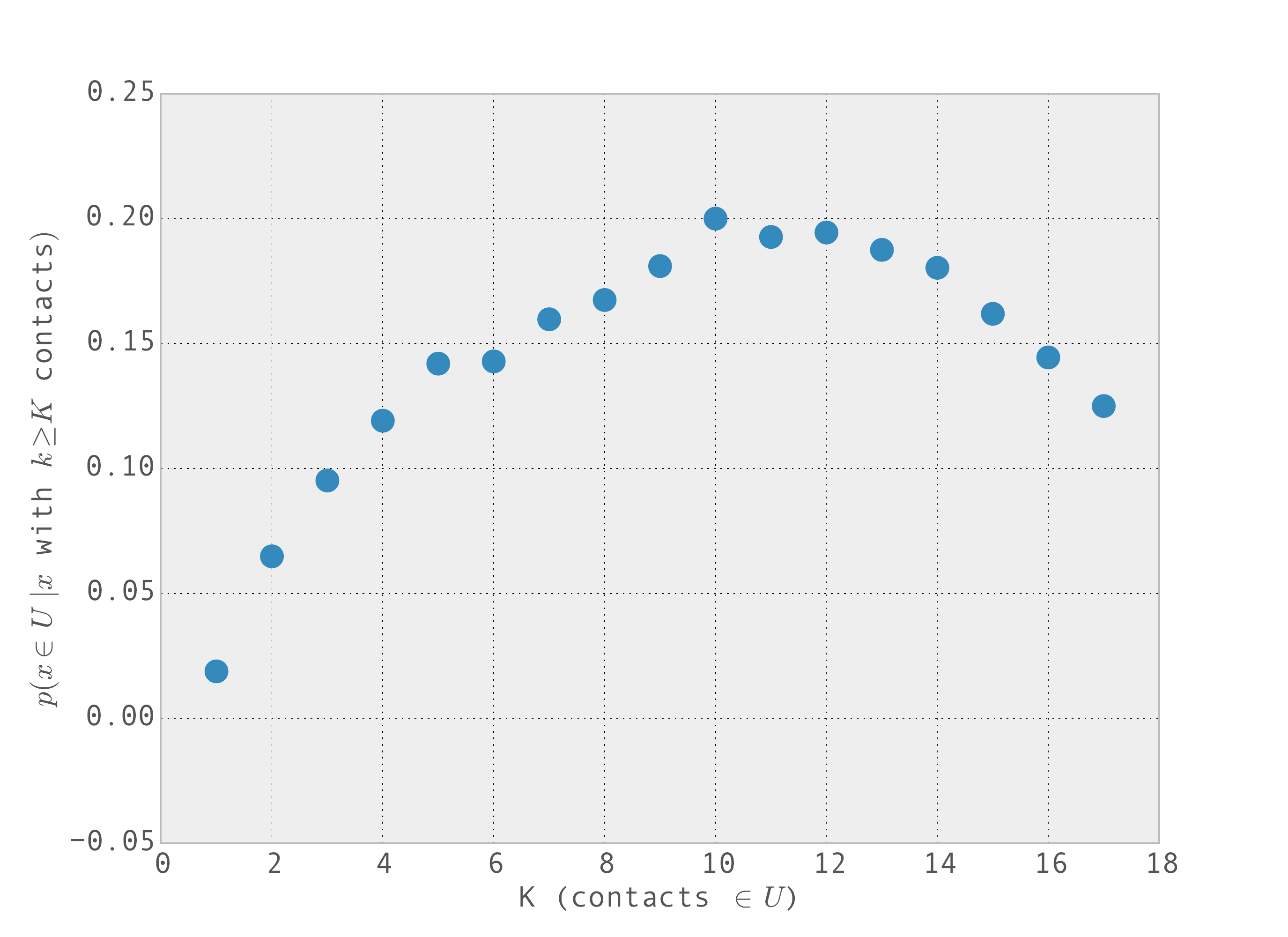}
	\vspace{-2em}
\end{center}
\caption{Probability of a user being in the event $U$, given the user has at least $K$ contacts in the event. 
}
	
\label{fig:PinEwamK}
\end{figure}

We also computed the ratio between the number of users in the event and the number of users with contacts in the event (such users can be at the event or not), both having at least $K$ contacts in the event $U$. 
This ratio is displayed in  
Fig.~\ref{fig:PinEwamK}, which  can be interpreted as the conditional probability that a user is in the event given that \textit{at least} $K$ of his/her contacts are in the event, as a function of $K$. 
This probability remarks the social character of the event because, as the number of contacts in the event increases, the probability of users with such contacts also increases up to $K=10$. 
Then, it starts to decrease due to the less significant proportion of people with more than $10$ contacts.
Notice that the sample space, where each probability is taken, is defined independently for each $K$.    

\section*{Conclusion and Future Work}

In this work, we have shown how we can combine the mobility and social network information present in mobile phone datasets to detect large social events and characterize their social features, such as an increase in local community structures of users present at an event. 
We have also shown that it is plausible to infer the probability of a user (in the contact graph) attending an event,
given the number of his/her contacts who attended the event. 

Finding useful inference algorithms of users attendance to a particular location given the mobility of their contacts can be of tremendous use for mobile phone companies. Mobile phone carriers have no mobility information about users in the contact graph $\calG$ who are not clients of the carrier. Therefore, having the possibility of making at least weak inferences about the locations and mobility of these users can add a valuable dimension for the carrier to better understand these (non client) users. 

We have presented this work within the formalism of time-varying  graphs~(TVG). Even though this work has not yet explored the potential uses of the  TVG to analyse the dynamics of the graph structure, we believe this is a useful formalism to extend this work into further understanding the changes in social dynamic networks as one approaches large-scale events.

\section*{Acknowledgements}

Authors are grateful to the STIC-AmSud program. Brazilian authors thank CAPES, CNPq, FAPERJ, and FINEP for their support.

\bibliographystyle{abbrv}
\bibliography{../biblio/mobility}

\begin{thebibliography}{1}

\bibitem{Becker2013}
R.~Becker, R.~C\'{a}ceres, K.~Hanson, S.~Isaacman, J.~M. Loh, M.~Martonosi,
  J.~Rowland, S.~Urbanek, A.~Varshavsky, and C.~Volinsky.
\newblock Human mobility characterization from cellular network data.
\newblock {\em Communications of the ACM}, 56(1):74--82, Jan. 2013.

\bibitem{Calabrese:2014}
F.~Calabrese, L.~Ferrari, and V.~D. Blondel.
\newblock Urban sensing using mobile phone network data: A survey of research.
\newblock {\em ACM Computing Surveys}, 47(2):25:1--25:20, Nov. 2014.

\bibitem{chua2013automatic}
F.~C.~T. Chua and S.~Asur.
\newblock Automatic summarization of events from social media.
\newblock In {\em International AAAI Conference on Weblogs and Social Media --
  ICWSM}, 2013.

\bibitem{ponieman2013human}
N.~Ponieman, A.~Salles, and C.~Sarraute.
\newblock Human mobility and predictability enriched by social phenomena
  information.
\newblock In {\em Proceedings of the 2013 IEEE/ACM International Conference on
  Advances in Social Networks Analysis and Mining}, pages 1331--1336. ACM,
  2013.

\bibitem{ritter2012open}
A.~Ritter, O.~Etzioni, S.~Clark, et~al.
\newblock Open domain event extraction from twitter.
\newblock In {\em Proceedings of the 18th ACM SIGKDD international conference
  on Knowledge discovery and data mining}, pages 1104--1112. ACM, 2012.

\bibitem{Wehmuth2014}
K.~Wehmuth, A.~Ziviani, and E.~Fleury.
\newblock {A Unifying Model for Representing Time-Varying Graphs}.
\newblock Technical Report RR-8466, INRIA, Feb. 2014.

\end{thebibliography}


\end{document}